%Paper: hep-ph/9312263
%From: JPHALYO@wiswic.weizmann.ac.il
%Date: Sun, 12 Dec 1993 11:26:55 GMT
%Date (revised): Wed, 16 Feb 1994 17:20:40 GMT

% Printing instructions:
%       This paper needs the macro package phyzzx.tex
%
\input phyzzx
\tolerance=1000
\sequentialequations
\def\rl{\rightline}

\def\r#1{$\bf#1$}

\def\t1{{\tilde 1}}

\def\AEF{A.E. Faraggi}

\def\NPB#1#2#3{Nucl. Phys. B {\bf#1} (19#2) #3}
\def\PLB#1#2#3{Phys. Lett. B {\bf#1} (19#2) #3}

\def\IJMP#1#2#3{Int. J. Mod. Phys. A {\bf#1} (19#2) #3}

\def\l{\langle}
\def\r{\rangle}
\def\D{$D_{45}~$}
\def\bD{$\bar D_{45}~$}

\REF\GSW{M. Green, J. Schwarz and E. Witten,
Superstring Theory, 2 vols., Cambridge
University Press, 1987.}
\REF\FSU{I. Antoniadis, J. Ellis and D. V. Nanopoulos, \PLB{194}{87}{231};
I. Antoniadis, J. Ellis, J. Hagelin and D. V. Nanopoulos, \PLB{205}{88}{459};
\PLB{208}{88}{209}; J. L. Lopez and D. V. Nanopoulos, \PLB{268}{91}{359}.}
\REF\MOD{\AEF, \PLB{278}{92}{131}.}
\REF\SLM{\AEF, \PLB{274}{92}{47}; \NPB{387}{92}{289}.}
\REF\DSW{M. Dine, N. Seiberg and E. Witten, \NPB{289}{87}{585};
J.J. Atick, L.J. Dixon and A. Sen, \NPB{292}{87}{109};
S. Cecotti, S. Ferrara and M. Villasante, \IJMP{2}{87}{1839}.}
\REF\FFF{I. Antoniadis, C. Bachas, and C. Kounnas, \NPB{289}{87}{87};
I. Antoniadis and C. Bachas, \NPB{298}{88}{586};
H. Kawai, D.C. Lewellen, and S.H. Tye,
Phys. Rev. Lett. {\bf57} (1986) 1832;
Phys. Rev. D {\bf 34} (1986) 3794;
Nucl. Phys. B {\bf 288} (1987) 1.}
\REF\NRT{\AEF, \NPB{403}{93}{101}.}
\REF\UP{E. Halyo, WIS-93/98/OCT-PH, hep-ph 9311300.}
\REF\KLN{S. Kalara, J. Lopez and D.V. Nanopoulos, \PLB{245}{91}{421};
\NPB{353}{91}{650}.}
\REF\CKM{A. E. Faraggi and E. Halyo, \PLB{307}{93}{305}; WIS--93/34/MAR--PH.
to appear in Nucl. Phys. B.}
\REF\FM{\AEF, WIS--92/81/OCT--PH, to appear in Nucl. Phys. B.}
\REF\PDG{Review of Particle Properties, Phys. Rev. D {\bf 45}, Vol. 11B.}
\REF\NIR{Y. Nir, in ``CP Violation", Lectures presented in the $20^{th}$
Annual SLAC Summer Institute on Particle Physics, 1992 and references therein.}
\REF\LQ{V. D. Agnelopoulos, J. Ellis, H.Kowalski, D. V. Nanopoulos, N. D.
Tracas and F. Zwirner, \NPB{292}{87}{59}.}
\REF\MIR{M. Leurer, WIS-93/90/SEP-PH.}
\REF\GG{B. R. Greene, K. H. Kirklin, P. J. Miron and G. G. Ross, \NPB{292}{87}
{602} and references therein.}
\REF\FLQ{S. Kelly, J. L. Lopez and D. V. Nanopoulos, \PLB{278}{92}{140}.}
\REF\FLQM{S. Kelly, J. L. Lopez and D. V. Nanopoulos, \PLB{261}{91}{424}.}

\singlespace
\rl{WIS--93/114/DEC--PH}
\rl{\today}
\rl{T}
\pagenumber=0
\normalspace
\smallskip
\titlestyle{\bf{TeV Scale Leptoquarks as a Signature of Standard--like
Superstring Models}}
\smallskip
\author{Edi Halyo{\footnote\dag{e--mail address: jphalyo@weizmann.bitnet}}}
\smallskip
\centerline {Department of Particle Physics,}
\centerline {Weizmann Institute of Science}
\centerline {Rehovot 76100, Israel}
\vskip 6 cm
\titlestyle{\bf ABSTRACT}

We show that there can be TeV scale scalar and fermionic leptoquarks with
very weak Yukawa couplings in a generic standard--like
superstring model. Leptoquark--(down--like) quark mixing though present, is
not large enough to violate the unitarity bounds on the CKM matrix. The
constraints on leptoquark masses and
couplings from FCNCs are easily satisfied whereas those from
baryon number violation may cause problems. The leptoquarks of the model are
compared to the ones in $E_6$ Calabi--Yau and flipped $SU(5) \times
U(1)$ models.

\singlespace
\vskip 0.5cm
\endpage
\normalspace

\centerline{\bf 1. Introduction}

Superstring theories are [\GSW], to date, the most promising Planck scale
theories of particle physics. In spite of their successes, one of the many
drawbacks of realistic superstring models is their loose connection to TeV or
weak scale physics. TeV (or weak) scale signs or predictions of realistic
superstring models
are rare even though they can reproduce most of the known low--energy physics.
It is important to look for these signs or predictions either to make specific
superstring models more plausible or to rule them out.

In this letter, we show that, under certain conditions, there can be TeV
scale leptoquarks in a class of standard--like superstring models [\MOD,\SLM].
We find that these leptoquarks have very weak (i.e. $<10^{-3}$) Yukawa
couplings if at all. Supersymmetry (SUSY) constraints in the observable and
hidden sectors play an important role in these results. One of the leptoquarks
mixes with down--like quarks, but the mixing is small enough to satisfy the
unitarity constraints on the CKM matrix. Due to the small
leptoquark Yukawa couplings, constraints from flavor changing neutral currents
(FCNCs) on leptoquark masses are easily satisfied. Baryon number violation may
impose severe constraints on leptoquark masses unless the Yukawa couplings
to diquarks are absent up to very high orders.
We also compare these
leptoquarks with those that arise from $E_6$ Calabi--Yau [\GSW] and flipped
$SU(5) \times U(1)$ [\FSU] models and discuss their differences.

The standard--like superstring models that we consider have the following
properties [\MOD,\SLM]:

1. $N=1$ space--time SUSY.

2. A $SU(3)_C\times SU(2)_L\times {U(1)^n}\times$hidden gauge group.

3. Three generations of chiral fermions
and their superpartners, with the correct quantum numbers
under ${SU(3)_C\times SU(2)_L\times U(1)_Y}$.

4. Higgs doublets that can  produce realistic electro--weak symmetry breaking.

5. Anomaly cancellation, apart from a single ``anomalous" U(1)
which is  canceled by  application of the
Dine--Seiberg--Witten (DSW) mechanism [\DSW].

The superstring standard--like models are constructed in the four
dimensional free fermionic formulation [\FFF].
The models are generated by a basis of eight boundary condition vectors
for all world--sheet fermions [\MOD,\SLM].
The observable and hidden gauge groups after application
of the generalized GSO projections are $SU(3)_C\times U(1)_C\times
 SU(2)_L\times U(1)_L\times U(1)^6${\footnote*{$U(1)_C={3\over 2}U(1)_{B-L}$
and $U(1)_L=2U(1)_{T_{3_R}}$.}}
and $SU(5)_H\times SU(3)_H\times U(1)^2$, respectively.
The weak hypercharge is given by
$U(1)_Y={1\over 3}U(1)_C + {1\over 2}U(1)_L$ and has the standard $SO(10)$
embedding. The orthogonal
combination is given by $U(1)_{Z^\prime}= U(1)_C - U(1)_L$.
The models have six right--handed and six left--handed horizontal symmetries
$U(1)_{r_j}\times U(1)_{\ell_j}$ ($j=1,\ldots,6$),
which correspond to the right--moving and left--moving world--sheet
currents respectively.

A generic standard--like superstring model including
the complete massless spectrum with the quantum numbers and the cubic
superpotential were presented in Ref. [\MOD] and will not be repeated here.
The notation of Ref. [\MOD] is used throughout this letter.

\bigskip
\centerline{\bf 2. SUSY constraints}

In order to preserve SUSY at $M_P$, one has to satisfy a set of F and D
constraints.
The set of F and D constraints is given by the following equations:
$$\eqalignno{&D_A=\sum_k Q^A_k \vert \chi_k \vert^2={-g^2e^{\phi_D}
\over 192\pi^2}Tr(Q_A) {1\over{2\alpha^{\prime}}}&(1a) \cr
&D^{\prime j}=\sum_k Q^{\prime j}_k \vert \chi_k \vert^2=0 \qquad j=1
\ldots 5 &(1b) \cr
&D^j=\sum_k Q^j_k \vert \chi_k \vert^2=0 \qquad j=C,L,7,8 &(1c) \cr
&W={\partial W\over \partial \eta_i} =0 &(1d) \cr}$$
where $\chi_k$ and $\eta_i$ are the fields that do and do not get VEVs
respectively
and $Q^j_k$ are their charges. $(2\alpha^{\prime})^{-1}=g^2M_P^2/{8 \pi}=M^2
\sim (10^{18}~GeV)^2$ and
$W$ is the superpotential. From Eq. (1a) we see that, $SO(10)$
singlet scalars must get VEVs $\sim g^2M/4 \pi \sim M/25$ in order to preserve
SUSY at $M_P$.

The set of F constraints in the observable sector has been studied before
[\NRT]. One finds that SUSY requires $\l \Phi_{12} \r=\l \bar \Phi_{12}
\r= \l \xi_3 \r=0$ even though the number of fields is larger than the number
of constraints. Then, one is left with only three F constraints from the
observable sector [\NRT].
%$$\eqalignno{&\bar \Phi_{23} \Phi_{13}+\bar \Phi_i^+ \bar \Phi_i^- =0 &(3a)\cr
%            &\bar \Phi_{13} \Phi_{23}+ \Phi_i^+  \Phi_i^- =0 &(3b)\cr
%            &\Phi_{45} \bar \Phi_{45}+ \Phi_i^+ \bar \Phi_i^+ +\Phi_i^-
% \bar \Phi_i^- =0 &(3c)} $$
F constraints in the hidden sector which are derived from the cubic
superpotential have also been investigated recently [\UP].
These lead to conditions on hidden sector VEVs which are
%$$\eqalignno{&H_{19}H_{20}+H_{23}H_{24}+H_{25}H_{26}=0 &(4a)\cr
%            &H_{13}H_{14}+H_{17}H_{18}=0 &(4b)\cr
%            &{1\over 2}\xi_1H_{24}+\Phi_{23}H_{26}=0 &(4c)\cr
%            &{1\over 2}\xi_1H_{23}+\bar \Phi_{23}H_{25}=0 &(4d)\cr
%            &{1\over 2}\xi_1H_{26}+ \bar \Phi_{23}H_{24}=0 &(4e)\cr
%            &{1\over 2}\xi_1H_{25}+\Phi_{23}H_{23}=0 &(4f)\cr
%            &{1\over 2}\xi_2H_{13}={1\over 2}\xi_2H_{14}={1\over
%%2}\xi_2H_{17}=
%{1\over 2}\xi_2H_{18}=0  &(4g)} $$
particularly strong if one also requires realistic quark and lepton masses.
%(This means that $\xi_1$ and $\xi_2$ in addition to
%$\Phi_1^\pm,\bar \Phi_2^\pm$ must get VEVs [\NRT].)
Then, SUSY in the hidden sector (at $M_P$) imposes
%$\l H_{13} \r=\l H_{14} \r=\l H_{17} \r=\l H_{18} \r=0$.
%$$\l H_{13} \r=\l H_{14} \r=\l H_{17} \r=\l H_{18} \r=0 \eqno(5)$$
$\l H_i \r =0$ where $i=13, \ldots, 26$ in the notation of Ref. 3 (with at most
one pair among these having non--zero VEVs in special cases) [\UP].

Once SUSY is dynamically broken by the hidden sector condensates, the
VEVs which vanish above can become non--zero. For broken SUSY,
$\l F \r \sim M_{SUSY}^2$ and $m_{3/2} \sim M_{SUSY}^2/M < O(TeV)$, in order
to solve the hierarchy problem. For a light gravitino (and light squark and
slepton masses), i.e. $m_{3/2} \sim O(100~GeV)$,
we need $M_{SUSY} \sim 10^{10}~GeV$ or $\l F \r \sim 10^{20}~GeV$. As a result,
the VEVs which vanished due to SUSY can now be non--zero and up to $O(TeV)$.
Note that for a heavy gravitino with $m_{3/2} \sim O(TeV)$ these VEVs can be up
to O($10~TeV$).

\bigskip
\centerline{\bf 3. Leptoquarks of the model}

In the massless $b_1+b_2+\alpha+\beta+(S)$ sector of standard--like superstring
models there are two color triplet, electroweak singlet states, $D_{45}$
and $\bar D_{45}$ [\MOD,\SLM].
Under $SU(3)_C \times SU(2)_L \times U(1)_C \times U(1)_L$, $D_{45}$ and
$\bar D_{45}$ transform as $(3,1,-1,0)$ and $(\bar 3,1,1,0)$ respectively.
Since $Q_Y=Q_C/3+Q_L/2$ and $Q_{Z^\prime}=Q_C-Q_L$ we find that $Q_Y(D_{45})
=Q_{EM}(D_{45})=-1/3$ and $Q_{Z^\prime}(D_{45})=-1$ with $\bar D_{45}$ having
opposite charges. Another combination of $Q_C$ and $Q_L$ gives $Q_{B-L}=2Q_C/3$
which is a gauge symmetry in these models. Thus, $Q_{B-L}(D_{45})=-2/3$ and
$Q_{B-L}(\bar D_{45})=2/3$.
{}From all these quantum numbers we see that \D and \bD are
actually leptoquarks. Note that \D and \bD are
superfields and as a result there are two scalar and two fermionic leptoquarks
in this model.

In general one expects that \D  and \bD  get large masses (of $O(10^{17}~GeV)$)
at the level of the cubic superpotential. Even if this is not the case \D and
\bD can get large masses from higher order terms (i.e. $N>3$ terms) in the
superpotential and decouple from the low--energy spectrum. In fact, in the
standard--like model under consideration, there are potential mass terms
for \D and \bD at the cubic level [\MOD,\SLM]
$$W_{D,\bar D}={1\over 2}D_{45} \bar D_{45} \xi_3+{1\over 2} D_{45} H_{18}
H_{21}  \eqno(2)$$
where $H_{21}$ is a hidden sector state which is a $\bar 3$ of color,
$\xi_3$ and $H_{18}$ are singlets of $SU(3)_C \times SU(2)_L
\times U(1)_C \times U(1)_L$
. We see that due to the SUSY constraints in the observable and hidden sectors,
i.e. since $\l \xi_3 \r=\l H_{18} \r=0$, \D and \bD remain massless at the
cubic
level of the superpotential.

As noted earlier there may be higher order terms which give large masses to
\D and \bD. Higher order ($N>3$) non--renormalizable
contributions to the superpotential are obtained by calculating correlators
between vertex operators [\KLN]
$A_N\sim\langle V_1^fV_2^fV_3^b\cdot\cdot\cdot V_N^b\rangle $
where $V_i^f$ $(V_i^b)$ are the fermionic (bosonic) vertex operators
corresponding to different fields. The non--vanishing terms are obtained by
applying the rules of Ref. [\KLN]. First, since only $H_{23},H_{25}$ or
$H_{24},H_{26}$ can get VEVs due to the SUSY constraints in the hidden sector,
a mass term containing
$H_{21}$ is not possible to any order in $N$. Second, $H_{21}$ gets a large
mass from the term ${1\over 2} H_{21}H_{22} \xi_1 $ in the cubic
superpotential and decouples from the low--energy spectrum [\MOD,\SLM].
Therefore \D cannot mix with $H_{21}$ at low or imtermediate energies.
On the other hand, there
are \D \bD mass terms which arise from $N>3$ terms in the superpotential. At
$N=5$ we find a large number of terms which can be combined to give
$$D_{45} \bar D_{45} (\xi_2+\xi_3){\partial W\over \partial \xi_3}+
\Phi_{12}{\partial W\over \partial \Phi_{12}}+\bar \Phi_{12}{\partial
W\over \partial \bar \Phi_{12}} \eqno(3)$$
These vanish because of the SUSY constraints, Eq. (2) and Eq. (3), which can
be written as
$${\partial W\over \partial \xi_3}={\partial W\over \partial \Phi_{12}}=
{\partial W\over \partial \bar \Phi_{12}}=0 \eqno(4)$$
There are many other higher order ($N>5$) \D \bD terms arising
from the observable states which are
proportional to the F terms in Eq. (8) and therefore vanish. There may be terms
which are not proportional to the F terms in Eq. (8) at very high orders. It is
difficult to disregard this possibility because the number of terms increases
rapidly with the order $N$. Here we assume that if there are such terms they
can be made to vanish by an appropriate choice of vanishing VEVs.

When hidden sector states are taken into account, there are $N=6$ terms
$$D_{45} \bar D_{45} T_2 \bar T_2 \Phi_{45} \Phi_2^+ (\xi_1+\xi_3)\eqno(5)$$
which give large masses to \D and \bD if $\Phi_2^+$ gets a VEV. Here $T_2, \bar
T_2$ are $5,\bar 5$ of the hidden $SU(5)_H$ gauge group. There are no
phenomenological constraints from quark and lepton masses or quark mixing on
$\l \Phi_2^+ \r$. In addition, the SUSY F and D constraints can be satisfied
whether $\l \Phi_2^+ \r$ vanishes or not. If $\l \Phi_2^+ \r \not=0$, then
generically
$\l \Phi_2^+ \r \sim M/10 \sim 10^{17}~GeV$ and the $\l T_2 \bar T_2 \r \sim
\Lambda_H^2$ where $\Lambda_H \sim 10^{14}~GeV$ is the hidden $SU(5)_H$
condensation scale [\CKM]. This gives $M_{D,\bar D} \sim 10^8~GeV$.
Potential leptoquark mass terms arising from VEVs of $H_i$ vanish due to the
SUSY constraints in the hidden sector, Eqs. (4).

If , on the other hand, $\l \Phi_2^+ \r=0$, then the \D \bD mass terms come
from the SUSY breaking VEVs. (Once again we assume that if there are $N>6$
terms similar to Eq. (9), then they can be made to vanish by an appropriate
choice of vanishing VEVs.) As mentioned earlier, the VEVs vanishing due to
SUSY can become non--zero (and up to the $TeV$ scale) once SUSY is broken.
Therefore, when SUSY is broken, \D and \bD get $TeV$ scale masses from the
cubic superpotential, i.e. from
the terms in Eq. (6) since now $\l \xi_3 \r \sim O(TeV)$ (for a scenario with
light squark and lepton masses). In
addition the scalar leptoquarks get contributions to
their masses from soft SUSY breaking terms. These are generally less than a
$TeV$. (For the case we consider, soft SUSY breaking masses for the scalars
$m_0\sim m_{3/2} \sim O(100~GeV)$.) Thus, under the conditions given above,
there are two scalar and two fermionic
leptoquarks with masses around the $TeV$ scale in this model.
The lower bound on $M_{D,\bar D}$ from direct leptoquark searches is $45~GeV$
[\PDG] which is easily satisfied.

The fermionic leptoquarks may mix with down--like quarks. In this model, there
is a $d_3^c D_{45}$ mixing term of the form
$$d_3^c D_{45} N_3 \Phi_{13} \Phi_3^+ (\xi_1+\xi_2+\xi_3) \eqno(6)$$
Similar mixing terms for the other two down--like quarks may appear at higher
orders, $N>6$. There are no $d_i \bar D_{45}$ mixing terms, for the
left--handed down--like quarks, due to the
conservation of $Q_L$ and $Q_C$. $\l N_i^c \r $ appears in
non--renormalizable terms which induce dimension four baryon number violating
operators [\NRT]. From the proton lifetime, we get the constraint $\l N_i^c \r
\sim O(TeV)$ at most. As a result, the mixing term in Eq. (10) is at most about
$O(GeV)$ and the others are smaller by at least an order of magnitude since
they appear at higher orders.
(Note that the mixing term in Eq. (10) can be made to vanish by taking
$\l \Phi_{13} \r=0$ or $\l \Phi_3^+ \r=0$.)
Now, the $3 \times 3$ CKM matrix becomes non--unitary
because of the new mixing terms in the $4 \times 4$ down quark mass matrix.
(In standard--like superstring models, CKM matrix arises mainly from the down
quark mass matrix [\CKM].) The strongest bounds on the magnitude of the
$d_3^c D_{45}$ mixing arise from unitarity of the CKM matrix ($V_{ij}$) which
imposes $|V_{uD}|<0.07$ [\PDG] and $|Re V_{id}^* V_{is}|<2.4 \times 10^{-5}$,
$i=u,c,t$ (from flavor changing $Z$ currents [\NIR]). In our case,
when $d_3^c D_{45}$ mixing is much smaller than $M_D$, $|V_{uD}|
\sim \l N_3^c \r/10^3 \l \xi_3 \r \sim 10^{-3}$ and $|V_{cD}| \sim |V_{tD}|$
are (at least) an order of magnitude smaller. With these results the
constraints from
unitarity are easily satisfied. Conversely, since the $d_3^c D_{45}$ mixing is
very small compared to $M_D$, there will not be an appreciable violation of
unitarity in the $3 \times 3$ CKM matrix.

\bigskip
\centerline{\bf 4. Leptoquark interactions}

The leptoquarks \D and \bD carry color, electric and $Z^{\prime}$ charge
and therefore have strong, electromegnetic and $Z^{\prime}$ gauge interactions.
Of these, the $Z^{\prime}$ interactions will be very weak at the $TeV$ scale if
the $Z^{\prime}$ gauge boson has a large mass (i.e $M_{Z^{\prime}}>>TeV$).
Otherwise all gauge interactions of \D and \bD are appreciable at the $TeV$
scale. In any case, the leptoquarks will be easy to produce in $e^+e^-$ or
$pp$ collisions. Production of leptoquarks has been investigated in Ref. [\LQ]
in detail. As final states \D and \bD will look like new, very massive,
$SU(2)_L$ singlet, down--like quarks.

Yukawa couplings of \D and \bD are more interesting since it is these that
allow \D and \bD to decay into quarks and leptons. In addition, Yukawa
couplings are model dependent and therefore useful to distinguish beteween
different models. The Yukawa couplings
allowed by $Q_C$ and $Q_L$ conservation are
$$W_Y=L_i Q_i \bar D_{45} \Phi+ e_i u_i D_{45} \Phi+N_i d_i D_{45} \Phi
\eqno(7) $$
where $i$ is the generation index ($i=3$ is the lightest generation in the
notation of Ref. [\MOD].) and $\Phi$ is a generic string of $SO(10)$
singlet fields which get VEVs. Effective Yukawa couplings for the three
terms ($g_{1i},g_{2i},g_{3i}$) are obtained
from the VEVs of the string of singlets divided by the proper power of $M$.
Each term in (11) also has a coefficient which can be calculated exactly and
is $O(1)$ [\KLN].

We look for terms which induce effective Yukawa couplings for \D and \bD at
orders $N>3$. We find that all kinds of couplings given in Eq. (11) are allowed
for all generations by the gauge symmetries of the model at $N=4$ and $N=5$.
All of these terms except one vanish due to the string selection rules as given
in Ref. [\KLN]. These selection rules arise from the left--handed $U(1)$
symmetries
and the world--sheet sigma model operators that appear in the vertex operators
in the world--sheet correlators (after picture changing is taken into account).
The term that remains at $N=5$ is
$$W_Y=L_3Q_3 \bar D_{45} \Phi_{45} \xi_3 \eqno(8)$$
which potentially gives an effective Yukawa coupling of $\l \Phi_{45} \xi_3 \r
/M \sim 10^{-2}$. But $\l \xi_3\r=0$ due to SUSY, so this term vanishes. (Even
after SUSY breaking, $\l \xi_3 \r/M \sim 10^{-15}$ and this term is
negligible.)
We see that the \D and \bD Yukawa couplings can only arise from terms at $N>5$
in this model. As a result they are at most $\sim 10^{-3}$ and probably
smaller which means that \D and \bD have very weak decays into leptons and
quarks.

If the Yukawa couplings of \D or \bD are not ``diagonal", i.e. \D or \bD
couple to more than one quark and lepton generation, they induce FCNCs.
FCNC processes such as $K_L \to e^+e^-$ and
$K^+ \to \pi^+ \nu \bar \nu$ give the strongest
bounds on $M_{D,\bar D}$ and the Yukawa couplings $g_{\{1,2,3\}i}$. From the
analysis of Ref. [\MIR] we get
$$|g_{j2}g_{j3}|<5.65 \times 10^{-8} sin \theta_c M^2_{D,\bar D} \eqno(9)$$
from $K_L \to e^+e^-$, and a slightly lower bound from
%$$|g_{j2}g_{j3}|<4.43 \times 10^{-8} sin \theta_c M^2_{D,\bar D} \eqno(14)$$
from $K^+ \to \pi^+ \nu \bar \nu$.
Here $j$ is the index for the different couplings
and $\theta_c$ is the Cabibbo angle with $sin \theta_c \sim 0.2$.

With the upper bound of $\sim 10^{-3}$ that we obtained for the \D and \bD
Yukawa couplings above, we find that the lower bound on $M_{D,\bar D}$ from
FCNCs is $M_{D,\bar D}>10~GeV$ which is not a constraint at all since the bound
from direct searches is $M_{D,\bar D}>45~GeV$. FCNC constraints are severe only
for leptoquarks with Yukawa couplings $\sim O(1)$ to more than one generation.
The same bounds and remarks also apply to the squark--quark--fermionic
leptoquark Yukawa couplings but since squark masses are very large (and
unknown) no useful bounds exist in this case.

\D and \bD can also have diquark couplings such as
$$W^{\prime}=u_i^c d_i^c \bar D_{45} \Phi + Q_i Q_i D_{45} \Phi \eqno(10)$$
(We will call these effective couplings $g_{4i}$ and $g_{5i}$.)
Baryon number (B) violation imposes
severe constraints on the strength of these couplings if there are non--zero
leptoquark couplings like those in Eq. (11). This is because if both leptoquark
and diquark couplings are non--zero at the same time, \D and \bD exchange leads
to very large B violating processes such as $Q_iQ_i \to u_je_j$ or $u_id_i
\to Q_jL_j$ where $i,j$ are generation indices.

Both kinds of diquark couplings for all generations are allowed by the local
symmetries of the model at $N=5$ and $N=6$. As for the leptoquark couplings,
all these terms except
one vanish due to the string selection rules. The term that remains at $N=5$ is
$$u_3^c d_3^c \bar D_{45} \Phi_{45} \xi_3 \eqno(11)$$
which vanishes due to SUSY ($\l \xi_3 \r=0$). Proton lifetime constrains the
product of leptoquark and diquark couplings such that
$${|g_{1i} g_{4i}|\over {M_{D,\bar D}^2}}<10^{-32}~GeV^{-2} \eqno(12)$$
or $|g_{1i} g_{4i}|<10^{-26}$ for $M_{D, \bar D} \sim O(TeV)$ with a similar
bound on $|g_{2i} g_{5i}|$. This requires a
search up to $N=11$ for both kinds of terms (and to higher orders for one if
the
other apperars at $N<11$) which is difficult to do due to the very large number
of terms at these orders. As before, we assume that if unwanted terms appear,
they can be made to vanish by an appropriate choice of vanishing VEVs. If this
cannot be done, one has to give very large masses to \D and \bD in order to
satisfy the constraint, Eq.(16), from B violation.

\bigskip\centerline{\bf 5. Discussion and Conclusions}

We found that, under certain conditions, standard--like superstring models
have two scalar and two fermionic leptoquarks with $M_{D,\bar D} \sim O(TeV)$
and the quantum numbers given above. $M_{D,\bar D} \sim O(TeV)$ only for light
($\sim O(100~GeV$)) squarks and sleptons. Moreover, \D and \bD have Yukawa
couplings to leptoquarks which are weaker than $\sim 10^{-3}$.
As we saw, SUSY constraints in the observable and hidden sectors play an
important role in these results. $D_{45}$ also mixes with the down quark (and
possibly with s and b). These mixings are small enough (compared to $M_D$) to
satisfy the unitarity constraints on the CKM matrix. FCNC constraints on the
leptoquark masses
can be easily satisfied due to the small Yukawa couplings. On the other hand,
baryon number violation may impose severe constraints on $M_{D,\bar D}$
if there are Yukawa couplings to both leptoquarks and diquarks at low $N$.

Leptoquarks also appear in $E_6$ Calabi--Yau (CY) and flipped $SU(5) \times
U(1)$ models. We now compare these with the leptoquarks of standard--like model
discussed above. In CY models,
leptoquarks are in each 27 of $E_6$, i.e. there is a leptoquark pair for
each generation. In the $SU(3)^3$ CY model [\GG], leptoquarks
get masses at $\sim 10^{12-14}~GeV$ where the gauge symmetry is broken
spontaneously and decouple from the spectrum. In CY models with a rank 5 gauge
group (e.g. $SU(3) \times SU(2) \times U(1)^2$ [\LQ]) leptoquarks are light
because the superpotential contains (for each generation)
$$W=\lambda_1 H_1 H_2 N +\lambda_2 D \bar D N \eqno(13)$$
in addition to leptoquark and diquark couplings of the form given in Eqs. (11)
and (14). Here $N$ is a $SO(10)$ singlet whose VEV gives the Higgs mixing as
well as the
leptoquark masses. Since Higgs mixing has to be $<O(TeV)$ in order to get
weak symmetry breaking, $M_{D,\bar D}<O(TeV)$ as in our case. Note that in the
CY model the scale of leptoquark masses is correlated with the Higgs mixing.
If Higgs mixing is small, then $M_{D,\bar D}$ can be smaller than $O(TeV)$.
There is no such connection between Higgs mixing and leptoquark masses in
standard--like models. Instead, the correlation is between the scale of the
sparticle spectrum and leptoquark masses via the gravitino mass.

In CY models, either the leptoquark or the diquark Yukawa couplings (or both)
are eliminated by the discrete symmetries $(-1)^{3B}$ or $(-1)^L$ respectively.
(B and L are baryon and lepton numbers respectively.) The one which is not
eliminated is, in general, $\sim O(1)$. In our case, the Yukawa couplings are
at most $\sim 10^{-3}$
and probably smaller. Also the discrete symmetries of CY models do not exist
in standard--like models since B and L are not good quantum numbers but only
(local) $B-L$ is. The explicit terms in Eqs. (12) and (15) are
counterexamples to these discrete symmetries.

Flipped $SU(5) \times U(1)$ models [\FSU] must also have leptoquarks (called
vector--like heavy quarks) in 5 or 10 representations of $SU(5)$ for the
gauge coupling unification scale to be
about the string unification scale, $10^{18}~GeV$ [\FLQ]. In the minimal case,
only one pair of these is needed but cases with more than one pair are also
possible. The leptoquark mass, in this case, can be computed from the
requirement of gauge coupling unification and is given by [\FLQ]
$$m_D=12.4 \left({TeV \over {\tilde m}} \right)^{37/12} \times (11.04^{\pm1})
GeV \eqno(14)$$
where $\tilde m$ is the gaugino or squark mass. Thus, $m_D$ can easily be
around or less than the $TeV$ scale as in our case. It has been
noted that, in these models, squark masses decrease as the leptoquark masses
increase which is exactly the opposite of what happens in standard--like
models. This point may be instrumental in distinguishing between them.
Much cannot be said about (leptoquark induced) B and L violation in the
flipped $SU(5)$ case since the Yukawa couplings of leptoquarks have not been
calculated.

If $TeV$ scale leptoquarks are observed, one can think of a number of scenarios
in which it would be possible to distinguish between the different superstring
models. For example if a) more than one pair of leptoquarks or b) leptoquarks
with Yukawa coupligs of $O(1)$ or c) sparticle masses much larger
than $O(100~GeV)$ are observed, standard--like models will probably be ruled
out. If no leptoquarks are observed at the $TeV$ scale, both
standard--like models and $SU(3)^3$ CY models are possible
but the rank 5 CY and flipped $SU(5)$ models are not. Finally, if the
amount of Higgs mixing turns out to be different than leptoquark masses
rank 5 CY models are ruled out. Standard--like and flipped $SU(5)$
leptoquarks are very similar to each other. It seems that the only way to
distinguish between them, until Yukawa couplings of the latter are known
in more detail, is by considering the sparticle mass scales in these models.
Then one can use the correlation (anti--correlation) between sparticle and
leptoquark masses as a possible signature of standard--like (flipped $SU(5)$)
models.

The leptoquarks of standard--like models are also interesting because of
their mixing with the down--like quarks. Since the CKM matrix is determined
mainly by $M_d$ in these models [\CKM], one may try to obtain the quark
mixing only from these terms.
In fact, this idea has been explored in flipped $SU(5)$ models in a qualitative
manner [\FLQM]. In addition,
this mixing may realize the Nelson--Barr mechanism which is a possible
solution to the strong CP problem, naturally. These issues
are currently under investigation and will be reported in the future.

\bigskip
\centerline{\bf Acknowledgements}
  I would like to thank Miriam Leurer for very useful discussions on
leptoquarks. This work is supported by a Feinberg Fellowship and the Depertment
of Physics.

\vfill
\eject

\refout
\vfill
\eject

\end
\bye